\documentstyle[12pt,epsfig]{article}
\def\bge{\begin{equation}}
\def\ene{\end{equation}}
\def\bg{\begin{eqnarray}}
\def\en{\end{eqnarray}}

\def\vr{\vec{r}}

\def\ageq{\stackrel{>}{\sim}}
\begin{document}
\renewcommand{\thefootnote}{\fnsymbol{footnote}}
\begin{flushright}
ADP-98-35/T308
\end{flushright}
%
%
\begin{center}
\begin{LARGE}
$\omega$-nucleus bound states in the Walecka model
\end{LARGE}
\end{center}
\vspace{0.5cm}
\begin{center}
{\large K.~Saito~\footnote{ksaito@nucl.phys.tohoku.ac.jp}} \\
Physics Division, Tohoku College of Pharmacy \\ 
Sendai 981-8558, Japan \\
{\large K.~Tsushima~\footnote{ktsushim@physics.adelaide.edu.au},} 
{\large D.H.~Lu~\footnote{dlu@physics.adelaide.edu.au} and }
{\large A.W.~Thomas~\footnote{athomas@physics.adelaide.edu.au}} \\
Special Research Center for the Subatomic Structure of Matter \\
and Department of Physics and Mathematical Physics \\
The University of Adelaide, SA 5005, Australia 
\end{center}
\vspace{0.5cm}
\begin{abstract}
Using the Walecka model, we investigate theoretically whether 
an $\omega$ meson is bound to finite nuclei. 
We study several nuclei from $^{6}$He to $^{208}$Pb, and 
compare the results with those in the quark-meson coupling (QMC) model. 
Our calculation shows that deeper $\omega$-nucleus bound states are 
predicted in the Walecka model than in QMC.  
One can expect to detect such bound states in the proposed 
experiment involving the (d,$^3$He) reaction at GSI.
\end{abstract}
PACS: 24.10.Jv, 21.10.Dr, 21.30.Fe, 14.40.-n 
%
%
\newpage

The study of the properties of hadrons in a hot and/or dense nuclear   
medium is one of the most exciting new directions 
in nuclear physics~\cite{qm97}--\cite{saitomega}.
The recent experimental data observed at the CERN/SPS by the 
CERES~\cite{ceres} and HELIOS~\cite{hel} collaborations has been 
interpreted as evidence for a downward shift of the
vector meson mass in dense nuclear matter~\cite{li}.
To draw a more definite conclusion,
measurements of the dilepton spectrum from vector
mesons produced in nuclei are planned at TJNAF~\cite{jlab} and
GSI~\cite{gsi} (see also Refs.\cite{ins,photo}).

Recently a novel approach to the study of meson mass 
shifts in nuclei
was suggested by Hayano {\it et al.}~\cite{hayano}, using the 
(d, $^3$He) reaction at GSI~\cite{pion} to produce real 
$\eta$ and $\omega$ mesons with nearly zero recoil.
If the meson feels a large enough, attractive (Lorentz scalar) force inside
a nucleus, the meson is expected to form meson-nucleus bound states.  
Hayano {\it et al.}~\cite{hayano2} have estimated the binding energies
for various $\eta$- and $\omega$-mesic nuclei.  We have also reported 
possibility of such bound states~\cite{etao} using the quark-meson coupling 
(QMC) model~\cite{saito}, in which the structure of the nucleus can be solved 
self-consistently, including the explicit quark structure of the nucleons.  
In this report we study several $\omega$-mesic nuclei  
($^{6}$He, $^{11}$B, $^{26}$Mg, $^{16}$O, $^{40}$Ca, $^{90}$Zr
and $^{208}$Pb -- the first three are the final nuclei in the 
proposed experiments at GSI~\cite{hayano,hayano2}) 
using an alternative, relativistic nuclear model, namely, the Walecka model 
or Quantum Hadrodynamics (QHD)~\cite{serot}. We compare the results with 
those found in QMC~\cite{etao}.  

\begin{figure}[hbt]
\begin{center}
\epsfig{file=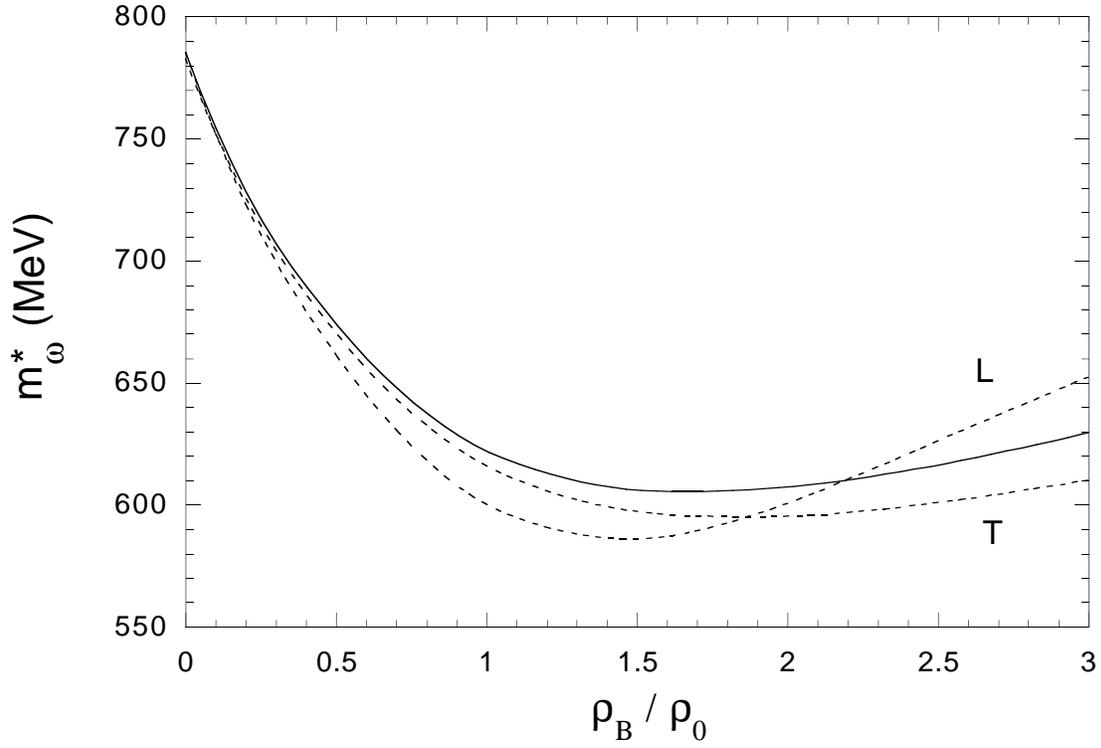,height=10cm}
\caption{The ``invariant'' mass of the $\omega$ meson in matter, 
including $\sigma$-$\omega$ mixing. 
The solid curve is for $q=1$ MeV, where the L and T modes are almost 
degenerate.  The dashed curves are for $q=500$ MeV, in which case 
the L and T modes are well separated. }
\label{fig:inv}
\end{center}
\end{figure}

In Ref.\cite{saitomega} we have already studied the propagation of the
$\omega$ meson with finite three momentum in 
infinite, symmetric nuclear matter within QHD-I,  
using the relativistic Hartree approximation.  
We also calculated the dispersion relation (in the time-like 
region) to get the ``invariant'' mass of the $\omega$ 
within the relativistic, random-phase approximation.  
The ``invariant'' mass, $m_\omega^*$, is defined by 
$\sqrt{q_0^2-q^2}$, where $q_0$ and $q=|{\vec q}|$ are the energy 
and three momentum of the $\omega$, respectively, and they are 
chosen so that the real part of the dielectric function in 
the full propagator vanishes. 
We do not repeat the details of the calculation here.  Instead, 
we show $m_\omega^*$ in Fig.~\ref{fig:inv} as a function 
of the nuclear density $\rho_B$. (For more information, see 
Ref.\cite{saitomega}.) 
The result shown includes the effect of $\sigma$-$\omega$ mixing 
in nuclear matter, which is, however, not large below normal 
nuclear matter density ($\rho_0=0.15$ fm$^{-3}$).  Furthermore, for 
low $q$ the separation between the longitudinal (L) and 
transverse (T) modes is very small.  

Since the proposed experiment at GSI~\cite{hayano,hayano2} might 
produce an $\omega$ meson with 
nearly zero recoil in a nucleus, it should be sufficient to consider 
the $\omega$ with low $q$ and ignore the separation between the L 
and T modes.  Using the results shown in the figure we 
shall parametrize the ``invariant'' mass of the $\omega$ with $q=1$ MeV 
(the solid curve in Fig.~\ref{fig:inv}) as a function of density.  
It is approximately given by
\bge
m_\omega^* \simeq m_\omega - 312.45 x + 199.40 x^2 - 59.277 x^3
    + 8.8427 x^4 - 0.52 x^5 , 
\label{param}
\ene
where all quoted numbers are in MeV, $m_\omega$(=783 MeV) 
is the mass in free space and $x=\rho_B/\rho_0$.  This reproduces 
$m_\omega^*$ well up to three times normal nuclear matter density. 

\begin{figure}[hbt]
\begin{center}
\epsfig{file=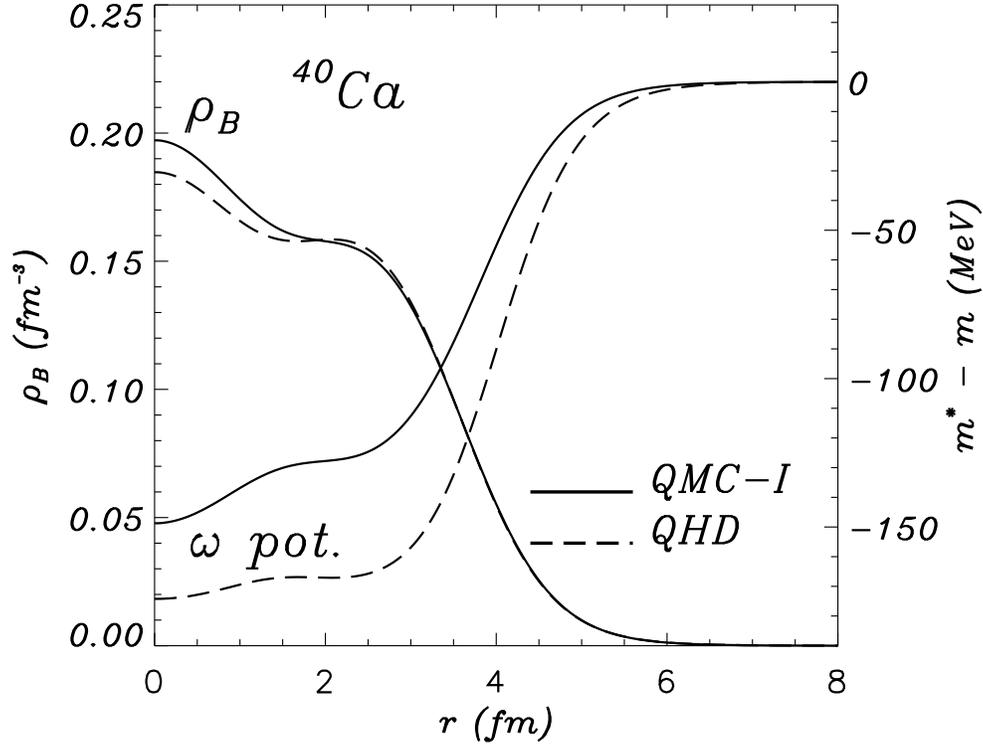,height=11cm}
\caption{
Potentials for the $\omega$ meson and the density distributions 
in $^{40}$Ca.  The results for QHD are shown by dashed curves, 
while those for QMC are shown by solid curves.  
}
\label{fig:pot40}
\end{center}
\end{figure}

Once one knows the density distribution of a nucleus, one can extract 
an effective potential for the $\omega$ meson from the ``invariant'' 
mass, assuming local density approximation.  Because the $\omega$ consists 
of the (same-flavor) quark and antiquark, we expect that the 
$\omega$ meson does not feel the repulsive,  
Lorentz vector potential generated by the nuclear environment.  
The total potential felt by the $\omega$ is then given by 
$m_\omega^*(\vr) - m_\omega$, 
where $m_\omega^*$ now depends on the position from the center of 
the nucleus.  In Fig.~\ref{fig:pot40}, we show the potential for 
an $\omega$ meson in $^{40}$Ca, together with the density distribution.  
We can see that the potential generated in QHD is rather deeper than 
that given by QMC. (To get the density distribution in QHD we have used 
the program of Horowitz {\it et al.}~\cite{horo}.)  
In a nucleus the (static) $\omega$-meson field, $\phi_\omega$, 
is then governed by the Klein-Gordon equation: 
\bge
\left[ \nabla^2 + E^2_\omega - m^{*2}_\omega(r) \right]\,
\phi_\omega(\vr) = 0 .
\label{kgeq}
\ene

An additional complication, which has not been added so far, is the 
meson absorption in the nucleus. This requires 
an imaginary part for the potential to describe the effect. 
At the moment, we have not been able to calculate the in-medium width of 
the meson, or the imaginary part of the potential in medium, 
self-consistently within the model. 
In order to make a more realistic estimate for the meson-nucleus bound states,
we shall include the width of the $\omega$ meson in the nucleus 
assuming the phenomenological form: 
\bg
\tilde{m}^*_\omega(r) &=&
m^*_\omega(r) - \frac{i}{2} 
\left[ (m_\omega - m^*_\omega(r)) 
\gamma_\omega + \Gamma_\omega \right], \\
\label{imaginary}
&\equiv& m^*_\omega(r) - \frac{i}{2} \Gamma^*_\omega (r),
\label{width}
\en
where $\Gamma_\omega$(=8.43 MeV) is the width in free space. 
In Eq.~(\ref{imaginary}), $\gamma_\omega$ is treated as a 
phenomenological 
parameter chosen so as to describe the in-medium meson width,  
$\Gamma^*_\omega$.

According to the estimates in Refs.~\cite{fri,kli}, 
the width of the $\omega$ at normal nuclear matter density 
is not large, typically a few tens of MeV: $\Gamma^*_\omega 
\sim 30 - 40$ MeV.
Thus, we calculate the single-particle energies using the values 
for the parameter appearing in Eq.~(\ref{imaginary}), 
$\gamma_\omega = 0, 0.2$ and $0.4$, which covers the estimated range. 
Thus we actually solve the following, modified Klein-Gordon equation:
\bge
\left[ 
\nabla^2 + E^2_\omega - \tilde{m}^{*2}_\omega(r) 
\right]\, \phi_\omega(\vr) = 0 .
\label{kgeq2}
\ene
Equation~(\ref{kgeq2}) has been solved in momentum 
space by the method developed in Ref.~\cite{landau}.
(We should mention that the advantage of solving the Klein-Gordon
equation in momentum space is that it can handle quadratic terms
arising in the potentials without any trouble, as
was demonstrated in Ref.~\cite{landau}.)

Now we are in a position to show our main results. 
In Tables~\ref{table:qhd} and \ref{table:qmc} the calculated single-particle 
energies for the $\omega$ meson are listed.  (In Table~\ref{table:qmc} 
the results of QMC~\cite{etao} are shown for comparison.)  
Our results suggest that one should expect to find 
bound $\omega$-nucleus states, as 
suggested by Hayano {\it et al.}~\cite{hayano,hayano2} and by our 
previous work~\cite{etao}. 
We have found that much deeper levels are predicted in QHD than in 
QMC because of the stronger, attractive force in QHD -- as shown 
in Fig.\ref{fig:pot40}. 
Note that the real part of the eigenenergy of the $\omega$ meson 
is very insensitive to the in-medium width.  We may understand this 
quantitatively, because the correction to the real part of 
the eigenenergy should be of order  
$\Gamma_\omega^{*2}/8m_\omega$, which is a few MeV (repulsive)
if we choose $\Gamma_\omega^* \sim$ 100 MeV. 
For a more consistent treatment, we need to calculate the in-medium meson 
width self-consistently within the model.

To summarize, we have calculated the single-particle energies for  
$\omega$-mesic nuclei using QHD and compared the results with those 
of QMC.   
Although the specific form for the width of the meson in medium  
could not be calculated in this model, 
our results suggest that one should observe $\omega$-nucleus bound 
states for a relatively wide range of the in-medium meson width.
In particular,  even in the light nuclei QHD gives very deep
single-particle levels ($\ageq$ 100 MeV), while QMC predicts 
much shallower levels.
If the $\omega$-nucleus bound states could be
observed in the future it would enable us to distinguish between
QHD and QMC.
%
%
\vspace{1.5cm}

\noindent{\bf Acknowledgment}\\
We would like to thank R.S Hayano, S. Hirenzaki, H. Toki and W. Weise 
for useful discussions. 
This work was supported by the Australian Research Council.
%
%
\newpage

%
%
\newpage
\begin{table}[htbp]
\begin{center}
\caption{
Calculated $\omega$ meson single-particle energies in QHD,
$E = Re(E_\omega - m_\omega)$,
and full widths, $\Gamma$, (both in MeV) in various nuclei, where
the complex eigenenergies are, $E_\omega = E + m_\omega - i \Gamma/2$.
See Eq.~(\protect\ref{imaginary}) for
the definition of $\gamma_\omega$. In the light of $\Gamma$ in
Refs.~\protect\cite{fri,kli}, the results with $\gamma_\omega = 0.2$ are
expected to correspond best with experiment.
The first three nuclei are the final nuclei in the proposed experiment
using the (d,$^3$He) reaction at GSI~\cite{hayano,hayano2}.
}
\label{table:qhd}
\begin{tabular}[t]{lc|cc|cc|cc}
\hline \hline
& &$\gamma_\omega$=0 & &$\gamma_\omega$=0.2& &$\gamma_\omega$=0.4& \\
\hline \hline
& &$E$ &$\Gamma$ &$E$ &$\Gamma$ &$E$ &$\Gamma$ \\
\hline
$^{6}_\omega$He &1s &-97.4&7.9 &-97.4&33.5 &-97.2&59.1 \\
\hline
$^{11}_\omega$B &1s &-129.0&8.0 &-129.0&38.5 &-128.9&69.0 \\
\hline
$^{26}_\omega$Mg &1s &-143.6&8.2 &-143.6&39.8 &-143.6&71.5 \\
                 &1p &-120.9&7.9 &-120.9&37.8 &-120.9&67.7 \\
                 &2s &-80.7&7.7 &-80.7&33.2 &-80.6&58.8 \\
\hline
$^{16}_\omega$O &1s &-134.1&8.1 &-134.1&38.7 &-134.0&69.3 \\
                &1p &-103.4&7.9 &-103.4&35.5 &-103.4&63.3 \\
\hline
$^{40}_\omega$Ca &1s &-147.6&8.2  &-147.6&40.1  &-147.6&72.0 \\
                 &1p &-128.7&8.0 &-128.6&38.3 &-128.6&68.6 \\
                 &2s &-99.8&7.8 &-99.8&35.6 &-99.8&63.5 \\
\hline
$^{90}_\omega$Zr &1s &-154.3&8.3  &-154.3&40.6  &-154.3&73.0 \\
                 &1p &-143.3&8.2  &-143.3&39.8  &-143.3&71.4 \\
                 &2s &-123.4&8.0 &-123.4&38.0 &-123.4&68.0 \\
\hline
$^{208}_\omega$Pb &1s &-157.4&8.4 &-157.4&40.8 &-157.4&73.3 \\
                  &1p &-151.3&8.3 &-151.3&40.5 &-151.3&72.7 \\
                  &2s &-139.4&8.1 &-139.4&39.5 &-139.4&70.8 \\
\hline \hline
\end{tabular}
\end{center}
\end{table}
%
%
\newpage
\begin{table}[htbp]
\begin{center}
\caption{
As in Fig.\protect\ref{table:qhd}, but for QMC. 
}
\label{table:qmc}
\begin{tabular}[t]{lc|cc|cc|cc}
\hline \hline
& &$\gamma_\omega$=0 & &$\gamma_\omega$=0.2& &$\gamma_\omega$=0.4& \\
\hline \hline
& &$E$ &$\Gamma$ &$E$ &$\Gamma$ &$E$ &$\Gamma$ \\
\hline
$^{6}_\omega$He &1s &-55.7&8.1 &-55.6&24.7 &-55.4&41.3 \\
\hline
$^{11}_\omega$B &1s &-80.8&8.1 &-80.8&28.8 &-80.6&49.5 \\
\hline
$^{26}_\omega$Mg &1s &-99.7&8.2 &-99.7&31.1 &-99.7&54.0 \\
                 &1p &-78.5&8.0 &-78.5&29.4 &-78.4&50.8 \\
                 &2s &-42.9&7.9 &-42.8&24.8 &-42.5&41.9 \\
\hline
$^{16}_\omega$O &1s &-93.5&8.1 &-93.4&30.6 &-93.4&53.1 \\
                &1p &-64.8&7.9 &-64.7&27.8 &-64.6&47.7 \\
\hline
$^{40}_\omega$Ca &1s &-111.3&8.2  &-111.3&33.1  &-111.3&58.1 \\
                 &1p &-90.8&8.1 &-90.8&31.0 &-90.7&54.0 \\
                 &2s &-65.6&7.9 &-65.5&28.9 &-65.4&49.9 \\
\hline
$^{90}_\omega$Zr &1s &-117.3&8.3  &-117.3&33.4  &-117.3&58.6 \\
                 &1p &-104.8&8.2  &-104.8&32.3  &-104.8&56.5 \\
                 &2s &-86.4&8.0 &-86.4&30.7 &-86.4&53.4 \\
\hline
$^{208}_\omega$Pb &1s &-118.5&8.4 &-118.4&33.1 &-118.4&57.8 \\
                  &1p &-111.3&8.3 &-111.3&32.5 &-111.3&56.8 \\
                  &2s &-100.2&8.2 &-100.2&31.7 &-100.2&55.3 \\
\hline \hline
\end{tabular}
\end{center}
\end{table}
\clearpage
%
\newpage
%
%
\newpage
%
%
\end{document}